\documentclass[preprint,amssymb,nofootinbib,amsmath,11pt]{JHEP3}

\JHEPspecialurl{http://jhep.sissa.it/JOURNAL/jhep3.tar.gz}

\usepackage{epsfig,multicol,amsmath}

\def\beq{\begin{equation}}

\def\eeq{\end{equation}}

\def\bea{\begin{eqnarray}}

\def\eea{\end{eqnarray}}

\def\eq#1{{Eq.~(\ref{#1})}}

\def\fig#1{{Fig.~\ref{#1}}}

\newcommand{\Lb}{\left(}

\newcommand{\Rb}{\right)}

\parskip=\itemsep               

\newcommand{\nn}{\nonumber}

\newcommand{\A}{{\cal A}}

\def\pom{{I\!\!P}}

\title{\LARGE \bf  N=4 SYM  model for soft interactions at high energy }

\author{\large  E. ~Levin$^{a,b}$\,\, and \,\,I.~Potashnikova$^{b}$  \\

a)  \,Department of Particle Physics, School of Physics and Astronomy,\\
Raymond and Beverly Sackler  Faculty
of Exact Science,  Tel Aviv University, Tel Aviv, 69978, Israel\\

b)\, Departamento de F\'{\i}sica, Centro de Estudios
Subat\'omicos, Universidad T\'ecnica Federico Santa Mar\'{\i}a,
and 
Centro Cient\'ifico-Tecnol\'ogico de Valpara\'iso,
Casilla 110-V, Valpara\'iso, Chile

}

\abstract{ In this paper we compare the prediction for high energy soft interactions in the model of $N=4$ SYM, with the experimental data. It is  
shown that this model is able to describe the total, elastic and inelastic cross sections and the elastic slope with only three free parameters. However, the model failed to obtain the cross sections for diffractive production, which was close to the experimental data, giving  small values for them. We believe that the theory of $N=4$ SYM, of the order of $1/\lambda$ is needed to find the origin of large mass diffraction.}

 \keywords{N=4 SYM, graviton reggeization,  eikonal approach}

\preprint{   TAUP \\

\today}

\begin{document}

\numberwithin{equation}{section}

\section{Introduction}

 It turns out that all soft interaction models  \cite{GLM,KAP,KMR,OST} have failed to predict the new LHC data (see Refs. \cite{ALICE,ATLAS,CMS,TOTEM}), showing that our understanding of long distance physics is  very limited (if any).
With this in mind it seems  reasonable to us,  to build a model for high energy strong interactions, based on $N=4$ SYM,
which is the only theory capable of treating long distance physics.
 $N=4$  SYM is a unique theory which allows us to study theoretically, the regime of the strong coupling constant  \cite{AdS-CFT}.

 Therefore in principle,  comparing the prediction for  the high energy scattering
amplitude in $N=4$ SYM with the experimental data,  we can single out the physics phenomena described by QCD more sucesfully than this simplfied approach.
It is well known (see for example  Refs.  \cite{POST,MHI,COCO,BEPI,LMKS} and references therein\footnote{We select here only those references which are 
related to the high energy scattering amplitude in $N=4$ SYM, which will be useful for an understanding of this paper.})  
that the main small parameters of $N=4$ SYM are

\beq \label{SPAR}
\lambda\,\,\gg\,\,1\,\Lb\rho\,\equiv \frac{2}{\sqrt{\lambda}}\,\ll\,1\Rb\,\,\,\,\mbox{but}\,\,\,\,g_S\,\ll\,1
\eeq

where 

\beq \label{PAR}
\lambda\,=\,4 \pi N_c g^2_{YM};\,\,\,g_s = \frac{g^2_{YM}}{4 \pi};\,\,L \,=\,\alpha'^{\frac{1}{2}}\,\lambda^{\frac{1}{4}};
\eeq
and $g_s$ is the string constant, $\alpha' \approx 1 \,GeV^{-2}$ is the slope of the reggeon trajectory, $N_c > 1$ is the number of colours,  and $L$ is the radius of AdS$_{5}$ space.

 As shown in our previous paper  \cite{LEPO1}, the theory of $N=4$ SYM in which $\rho $ is small (say $\leq 0.25$),  could be responsible for a small part of the total cross
 section, for energies below the LHC energy range.
In this paper we build an $N=4$ SYM model based on the  classical understanding of this word: a model is the theory which we apply  in the kinematic region
where the theory can't be proved to be incorrect. We expand the  theory of $N= 4$ SYM developed for $\rho \ll 1$, up to arbitrary values of $\rho$.

Comparing with the experimental data, we find the value of $\rho$ to be  $0.7 \div 0.8$.
We are able to describe the experimental data on the total, inelastic and elastic cross sections in the entire range of energies, including the LHC range.
However, we failed to describe the values and energy behaviour for diffractive production. We conclude from this failure, that we need to include high mass diffractive
 production which traditionally  stems from the triple Pomeron interaction. This interaction has not appeared in $N=4$ SYM for small $\rho$.  Therefore, we need to find
 the $\rho^2$-corrections,
in order to treat  diffractive production in $N=4$ SYM.

\section{Main formulae for high energy scattering amplitude in N=4 SYM}

\subsection{Reggeized graviton(Pomeron) propagator}

Before discussing the general approach  of  $N=4$ SYM to high energy scattering, we would like to draw the reader's  attention to the fact  that there exists
 two different regions of energy, that we have to consider in $N=4$ SYM: $\rho \alpha' s < 1$ and  $\rho \alpha' s < 1$ (see \eq{PAR}).
 In the first region, the multiparticle production has a very small cross section, and it can be neglected. However, in the second region the graviton reggeization
 leads to the inelastic cross section that is rather large, and at ultra high energies the scattering amplitude reveals all of the typical features of the black disc
regime: $\sigma_{el}  =  \sigma_{in} = \sigma_{tot}/2$.

At $\rho \to \infty$ the main contribution  to the scattering amplitude at high energy  in
$N=4$ SYM stems from the exchange of one graviton. The  formula for this exchange has been written in
Ref. \cite{BST2,COCO,LMKS}.
In $ AdS_5 =AdS_{d + 1}$ space this amplitude takes the following form:

\beq \label{1GE}
A_{1GE}(s,b; z,z')\,\,=\,\,g^2\,\,T_{\mu\nu}\Lb p_1,p_2\Rb G_{\mu \nu \mu^{\,\prime} \nu^{\,\prime}}\Lb u\Rb\,T_{\mu^{\,\prime}\nu^{\,\prime}}\Lb p_1,p_2\Rb\,\, \xrightarrow{s\gg \mu^2} g^2 \, s^2\, z^2\,z'^2 G_3\Lb u \Rb
\eeq

\FIGURE[h]{\begin{minipage}{50mm}

\centerline{\epsfig{file=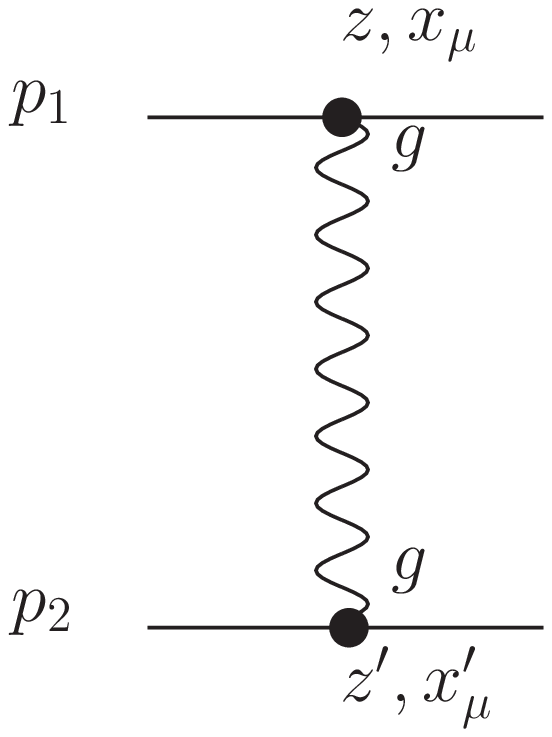,width=40mm}}

\caption{The one graviton (1GE) exchange.  }

\end{minipage}

\label{oge}}

where $T_{\mu\nu}$ is the energy-momentum tensor, and $G$ is the
propagator of the massless graviton. The last expression in
\eq{1GE}, reflects the fact that for high energies, $T_{\mu\nu} =
p_{1,\mu} p_{1,\nu}$  and at high energies the momentum transferred $q^2 \,\to q^2_{\perp}$ which led to  $G_3\Lb u\Rb $ (see  Refs. \cite{BST2,LMKS}).  In $AdS_5$  the metric takes  the following form

\beq \label{N44}
d s^2\,\,=\,\,\frac{L^2}{z^2}\,\Lb \,d z^2\,\,+\,\,\sum^d_{i=1}  d x^2_i \Rb\,=\,\frac{L^2}{z^2}\,\Lb \,d z^2\,+\,d \underline{x}^2 \Rb
\eeq
where $u$ is a new variable which is equal to

\beq \label{u}
u\,\,=\,\,\frac{ (z - z')^2 + (\underline{x} - \underline{x}')^2}{ 2 \,z\,z'}\,\,=\,\,\frac{ (z - z')^2 + b^2}{ 2 \,z\,z'}\,\,
\eeq

and

\beq \label{G}
G_{3}\Lb u \Rb \,\,=\,\,\frac{1}{4 \pi}\,\frac{1}{\left\{ 1 + u + \sqrt{u (u + 2)}\right\}^2\,\sqrt{u (u + 2)}}
\eeq

where $b$ is the impact parameter (see \fig{oge}). As one can see from \eq{1GE}, the one graviton exchange amplitude is real and increases as $s^2$. 
 However,  the high energy limit of \eq{1GE} has to be modified, since in the order of $\rho$, the intercept of the graviton decreases and becomes
equal to $2 - \rho$ (see Refs/ \cite{BST1,BST2,BST3,MHI,COCO}).  
The resulting expression for \eq{1GE} which we can consider to be the propagator of the  reggeized graviton exchange,  takes the following form:

\beq \label{POM}
A_{1GE}(s,b; z,z')\,\rightarrow\,\,P\Lb s,b; z, z'\Rb\,=\,\Lb \cot\Lb \pi \rho/2\Rb \,+\,i \Rb \frac{\xi_0}{\sinh\Lb \xi_0\Rb\,\tau^{3/2}\sqrt{u ( u + 2)}}\,\exp\Big(  ( 1 - \rho) \,\tau\,-\,\frac{\xi^2_0}{\rho \,\tau}\Big)
\eeq

  where

\beq \label{NOTA}
\tau\,=\,\ln\Lb \rho z\,z'\,s/2\Rb\,;\,\,\,\,\,\xi_0\,=\,\ln\Lb 1 \,+\,u\,+\,\sqrt{u (2 + u)}\Rb\,;\,\,\,\,\,
\eeq

One can see that \eq{POM} does not describe the exchange of the Regge pole with the intercept $(1 - \rho)$, but it is similar to the contribution of the BFKL Pomeron  which
is responsible for  high energy scattering in $N=4$ SYM,
 but at $g_s \ll 1$.
 Therefore, \eq{POM} gives an explicit example of AdS-CFT correspondence  \cite{AdS-CFT} which inherits all of the bad features of the BFKL Pomeron 

 \cite{BFKL}.

 Firstly, for large values of the impact parameters, \eq{POM} falls down only as a power of $b$, and such a decrease results  in a power -like increase of the radius of
 interaction, in the direction which contradicts the Froissart theorem  \cite{FROI} (see Ref. \cite{KOWI}). The second characteristic property of \eq{POM}, 
as well as the BFKL Pomeron,  is the absence of the shrinkage of the diffractive peak for high energy scattering.

  In principle, the shadowing corrections which should be large in our case since the amplitude of \eq{POM} increases with energy, can generate
 the effective shrinkage of the diffractive peak. However, we have decided to introduce  some corrections to \eq{POM}, namely,

\beq \label{POMMO}
P\Lb s,b; z, z'\Rb \,\,\longrightarrow\, \,\hat{P}\Lb s,b; z, z'\Rb\,\,\equiv\,\,P\Lb s,b; z, z'\Rb\,e^{  - \sqrt{2 \rho \alpha'}\,b\,\,-\,\,\frac{b^2}{2\,a\, \alpha' \rho\,\tau}}
\eeq

The first term in the exponent of \eq{POMMO} includes the minimal mass that appears in  string theory, which $N=4$ is the the limit at $\rho \ll 1$.
 The second one takes  the following form in $q$ representation: $ \exp\Lb- q^2\,\Lb a \rho/2\Rb \tau\Rb$ and corresponds to the shrinkage of the diffraction
 peak, with the effective $\alpha' = a \rho/2$.

\subsection{Vertices for Pomeron interaction with colliding hadrons}

It's well known that the contribution of the Pomeron to the scattering amplitude  takes the following form

\beq \label{POMAM}
A_{\pom}\Lb s, q; z, z'\Rb\,\,=\,\,g^2_{\pom}\Lb q\Rb\,\hat{P}\Lb s,q; z, z'\Rb\,\,=\,\,g^2_{\pom}\Lb q\Rb\,\int d^2 b e^{i \vec{q} \cdot \vec{b}}\,\hat{P}\Lb s,b; z, z'\Rb
\eeq

where $g(q)_\pom$ are the vertices of the Pomeron-hadron interaction. The vertex for the interaction of the graviton with the hadron has been found in the soft wall model
 for confinement  \cite{SOFTW}, that describes a number of different hadron characteristics  \cite{HADPROP}. It has the following form (see Ref. \cite{ABCA}):

\beq \label{VER1}
 g_\pom\Lb q_T; z \Rb\,\,=\,\,2\,g_{YM}\,H\Lb q_T, z\Rb \,\,=\,\,2 g_{YM}\,\Gamma\Lb a + 2\Rb\,U\Lb a, -1, 2 \xi\Rb\,\,=\,\,2 g_{YM}\,\,a (a + 1)\int^1_0 d t \,t^{a - 1}\, (1 - x)\, e^{- 2 \xi t/(1 - t)}
\eeq

where $U\Lb a, -1, 2 \xi\Rb$ is Tricomi's (confluent hypergeometric) function  \cite{MATH} and

\beq \label{PARVE}
\xi \,=\,\kappa^2 z^2;\,\,\,\,\,\,
\,a\,\,=\,\,\frac{q^2_T}{8 \kappa^2};\,\,\,q^2_T\,\,=\,\,- t  \leftarrow\,\mbox{momentum transferred by the Pomeron}
\eeq

where $\kappa$ is the dimensional parameter of the soft wall approach, and $\kappa^2 = m^2/8$ where $m$ is the nucleon mass
(see Ref. \cite{HADPROP} for more details). We replace the exact formula of \eq{VER1} by the following  approximate expression

\beq \label{VER2}
 g_\pom\Lb q_T; z \Rb\,\,=\,\,2\,g_{YM}\,\exp\Big( - B\Lb \xi\Rb\,a\Big)\,\,\,\,
\mbox{with}\,\,\,\, B\Lb \xi\Rb\,=\,1 \,-\, C \,+ \,U^{(1,0,0)}\Lb 0, -1, 2 \xi\Rb
\eeq

which allows us to perform analytically,  part of the intergrations. $C$ is the Euler constant in \eq{VER2}.

Finally, the amplitude for the exchange of a single Pomeron in impact parameter representation,

 which we need in order to take into account the eikonal rescattering, has the form

\beq \label{POMFAM}
\tilde{A}_\pom\Lb b, s; z, z'\Rb\,\,=\,\,\frac{1}{2 \bar{B}}\,e^{ - \frac{b^2}{4 \bar{B}}}\int d^2 b' I_0 \Lb \frac{\vec{b}\cdot \vec{b}'}{2 \bar{B}}\Rb\,\hat{P}\Lb s,b; z, z'\Rb
\eeq

\subsection{Eikonal formula}

\FIGURE[h]{\begin{minipage}{50mm}

\centerline{\epsfig{file=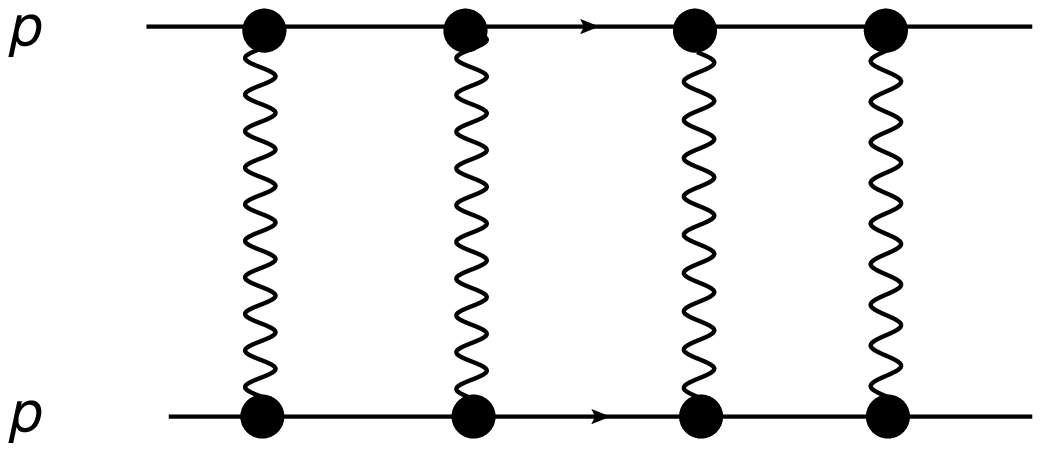,width=40mm}}

\caption{The eikonal formula that takes into account the multiPomeron exchanges.  }

\end{minipage}

\label{muge}}

As we have mentioned, $A_{\pom}\Lb s, q; z, z'\Rb$  increases steeply with energy $s$, and has to be unitarized using the eikonal formula 
 \cite{BST2,BST3,MHI,COCO,LMKS} (see \fig{muge})

\beq \label{EIK}
A_{eikonal}\Lb s,b;z,z'\Rb \,\,=\,\,i \Lb\,1\,\,\,-\,\,\,\exp
\Lb i\,  \tilde{ A}_{\pom}\Lb s, b;z, z'| \eq{POMMO}\Rb\Rb\Rb
 \eeq

In Ref. \cite{BST1,MHI} it was argued that AdS/CFT correspondence leads to the corrections to \eq{EIK} which are small
$(\propto 2/\sqrt{\lambda})$. The unitarity constraints for \eq{EIK} take the following form

\beq \label{EIKUN}
2\,\mbox{Im} \A_{eikonal}\Lb s,b; z,z'\Rb\,\,\,=\,\,\,| A_{eikonal}\Lb s,b; z,z'\Rb|^2\,\,\,+\,\,{\cal O}\Lb \frac{2}{\sqrt{\lambda}}\Rb
\eeq

\subsection{Nucleon-nucleon  scattering at high energy: observables}

We only need to integrate \eq{EIK} with the wave functions of the proton,
in order to obtain the amplitude for propton-proton scattering.
 Indeed the proton-proton amplitude is equal to

\beq \label{PPAM}
A_{pp}\Lb s,b\Rb\,\,=\,\,i\,\int d z\, d z' \Phi\Lb z\Rb \Phi\Lb z'\Rb \Lb\,1\,\,\,-\,\,\,\exp
\Lb i\,   A_{\pom}\Lb s, b; z, z'|\eq{POMMO}\Rb\Rb\Rb
\eeq

Based on  \eq{PPAM}, we can  calculate the number of the experimental observable.  We list below those of them that we actually use in our description.

\bea 
\sigma_{tot} &=&\,\,2 \int d^2 b \int d z\, d z'\, \Phi\Lb z\Rb \Phi\Lb z'\Rb 
\Big( 1 -  \cos\Lb \mbox{Re}A_{\pom}\Rb e^{ - \mbox{Im} A_\pom}\Big); \label{sitot}\\
\sigma_{el} &=& \int d^2 b \Lb  \Big\{\int d z\, d z'\, \Phi\Lb z\Rb \Phi\Lb z'\Rb\Big( 1 -  \cos\Lb \mbox{Re}A_{\pom}\Rb e^{ - \mbox{Im} A_\pom}\Big)\Big\}^2\right. \nn\\
& +& \left.  \Big\{\int d z\, d z' \,\Phi\Lb z\Rb \Phi\Lb z'\Rb\Big( \sin\Lb \mbox{Re}A_{\pom}\Rb e^{ - \mbox{Im} A_\pom}\Big)\Big\}^2\Rb;\label{sirl}\\
\sigma_{sd}\ &=& 2 \int d^2 b \int d z\, \Phi\Lb z\Rb\Big\{ \Lb \int d z' \,\Phi\Lb z'\Rb \Big( 1 -  \cos\Lb \mbox{Re}A_{\pom}\Rb e^{ - \mbox{Im} A_\pom}\Big)\Rb^2 \,\nn\\
&+&\,\Lb \int d z'\, \Phi\Lb z'\Rb \Big( \sin\Lb \mbox{Re}A_{\pom}\Rb e^{ - \mbox{Im} A_\pom}\Big)\Rb^2\Big\}\label{sisd}\\
\sigma_{dd} &=& \int d^2 b \int d z\, d z' \,\Phi\Lb z\Rb \Phi\Lb z'\Rb \Big\{\Big( 1 -  \cos\Lb \mbox{Re}A_{\pom}\Rb e^{ - \mbox{Im} A_\pom}\Big)^2 + \Big( \sin\Lb \mbox{Re}A_{\pom}\Rb e^{ - \mbox{Im} A_\pom}\Big)\Big\}\label{sidd}\\
B_{el} &=& \frac{\int b^2 d^b \int d z\, d z' \,\Phi\Lb z\Rb \Phi\Lb z'\Rb
\Big( 1 -  \cos\Lb \mbox{Re}A_{\pom}\Rb e^{ - \mbox{Im} A_\pom}\Big)}{2 \int d^2 b \int d z\, d z' \Phi\Lb z\Rb \Phi\Lb z'\Rb
\Big( 1 -  \cos\Lb \mbox{Re}A_{\pom}\Rb e^{ - \mbox{Im} A_\pom}\Big)}\label{bel}
\eea

\subsection{Proton wave function}

The last ingredient of our approach is the wave function of the proton which we chose in the soft wall approximation  \cite{SOFTW}. It is given by:

\beq \label{WAVEF}
\Phi\Lb \xi_i\Rb\,\,=\,\,\kappa \Big( \frac{1}{\Gamma\Lb 5/2\Rb}\xi^4_i\,\,+\,\,\frac{1}{\Gamma\Lb 7/2\Rb}\xi^6_i\Big)\,e^{ - \xi^2_i}
\eeq

This wave function works quite well for describing the key properties of the hadron, as  shown in Refs. \cite{HADPROP}.
It is easy to see that

\beq \label{WAVEFN}
\int\,d z \,\Phi\Lb z \Rb \,\,=\,\,1
\eeq

\section{Description of  the experimental data}

\subsection{Fitting parameters}

As one can see our main formulae (see \eq{POMMO},\eq{POMAM},\eq{VER2} and \eq{EIK})  contain a number of parameters.
We have fixed all of them except for three, from the description of the hadron characteristics in the soft wall model (see Refs.  \cite{SOFTW,HADPROP}. These three fitting parameters are

\beq \label{PARFIT}
g_s\,=\,\frac{g^2_{YM}}{4 \pi};\,\,\rho = 2/\sqrt{\lambda};\,\,\,\,a
\eeq

In $N=4$ SYM with AdS/CFT correspondence, $g_s$ and $\rho$ are much less than  1 while $a$ =0.
In our model we consider them as free parameters that can have arbitrary values.
 The difference between the values of these parameters found from comparing our formulae with the experimental data,
 and the  $N=4$ SYM expectations, will tell us how the $N=4$ SYM model differs from the theory.
In principle, using the actual values of these three parameters, we can also estimate up to what order of magnitude in $\rho^n$ we need to expand,
 in order to calculate the scattering amplitude, and 
to obtain a description of the experimental data.

\subsection{The result of the fit}

Using \eq{POMMO}, \eq{POMAM}, \eq{VER2} and \eq{EIK} we fit the experimental data starting with $W=\sqrt{s} =20\, GeV$. We do not include the LHC data in the fit 
since we wish to check the predictive power of the model.  \fig{sig} and \fig{bel} show the best result of our fit for $\sigma_{tot}$, 
$\sigma_{el}$ and the slope for $d \sigma_{el}/ d t $ at $t=0$.  The values of the parameters are $g_s = 0.245 \pm 0.003$, $\rho = 0.797 \pm 0.001$ and $a = 0.23$ $ \pm$ 0.005 $GeV^{-2}$.  The value of $\chi^2/d.o.f.  =  1.25$, which is very good.

\FIGURE[h]{

\centerline{\epsfig{file=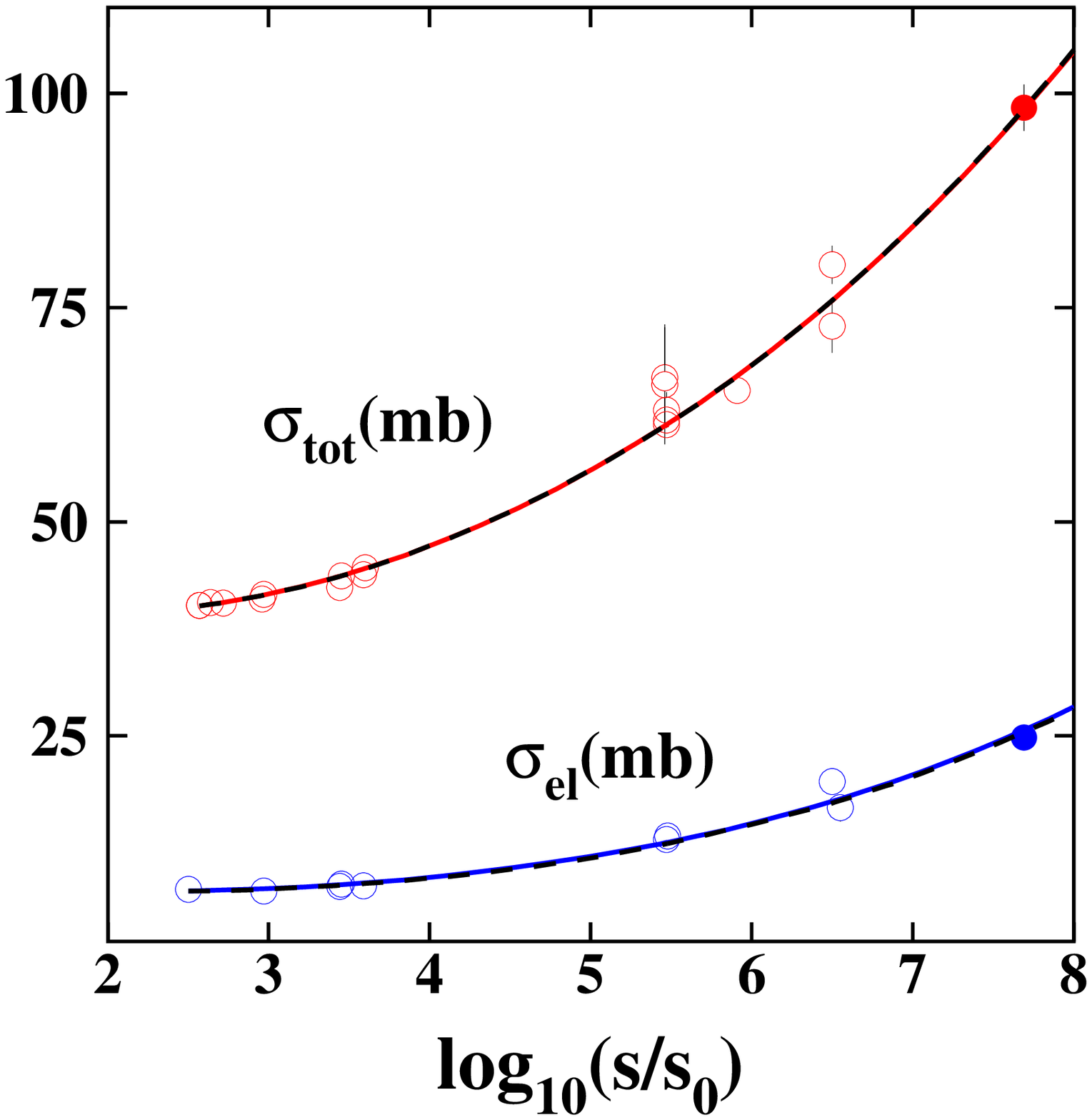,width=100mm,height=80mm}}

\caption{The description of the energy behaviour of the total and  elastic cross sections  with  $g_s = 0.245$, $\rho = 0.797$ and $a = 0.23$ $GeV^{-2}$. 
The solid line corresponds to the fit without the LHC data, while the dashed line describes the result of the fit that includes the LHC data.
The open circles show the data from PDG \cite{PDG}, and the full circls show the LHC data \cite{TOTEM}.}
\label{sig} }

~

~

\FIGURE[h]{
\centerline{\epsfig{file=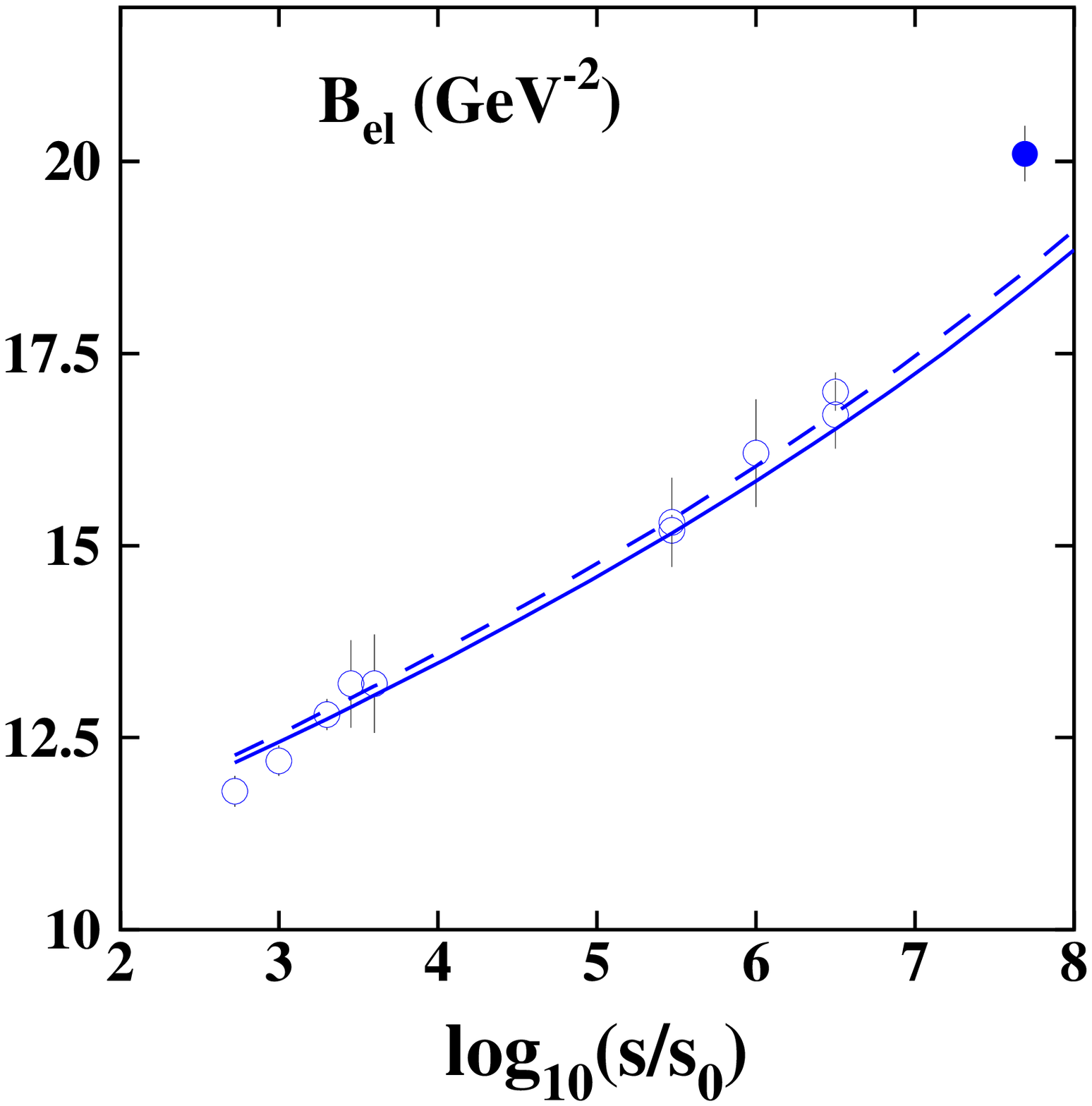,width=100mm,height=80mm}}
\caption{The description of the energy behaviour of the slope for the   elastic cross section,  with the same values of the parameters as in \fig{sig}. 
The solid line corresponds to the fit without the LHC data, while the dashed line describes the result of the fit that includes the LHC data.
 The open circles show the data from PDG  \cite{PDG}, and the full cuircles show the LHC data  \cite{TOTEM}.}
\label{bel} }

However, when we include the data for  the diffractive cross section for  either  the single diffraction or the double diffraction, we failed to describe these sets of 
data. Our best values for these cross sections turn out to be less than 2 mb.
Such values are in striking contradiction with the experimental data.
 For example at $W=7\, TeV$, the values for the single diffractive cross section are as large as 14 mb. The reason why this happens is clear from \fig{am}.
 In these figures, we show the behaviour of $\Phi\Lb z \Rb$, as well as the real and imaginary parts of the scattering amplitude, as functions of $z$.

 One can see that the amplitude depends only mildly on $z$ in the region where $\Phi\Lb z \Rb$ is not very small.
Therefore, in \eq{sitot}-\eq{bel} we can take the term for the amplitude outside of the integral at $z=z_{max}$, where $z_{max}$ is the value of $z$ where the function
 $\Phi\Lb z \Rb$ has a maximum. In doing so, we obtain the result that the integral over $z$ is equal to 1, and the cross sections for diffractive processes
 are equal to zero. At large $b$, the situation changes and the amplitude starts to depend on $z$ in the region where $\Phi\Lb z \Rb$ is not small.
 In this case we can see the difference in the integrals for elastic, single and double diffraction.  Unfortunately, the region of large $b$ gives only a small
 contribution.

\FIGURE[h]{
\begin{tabular}{c c c}
\epsfig{file=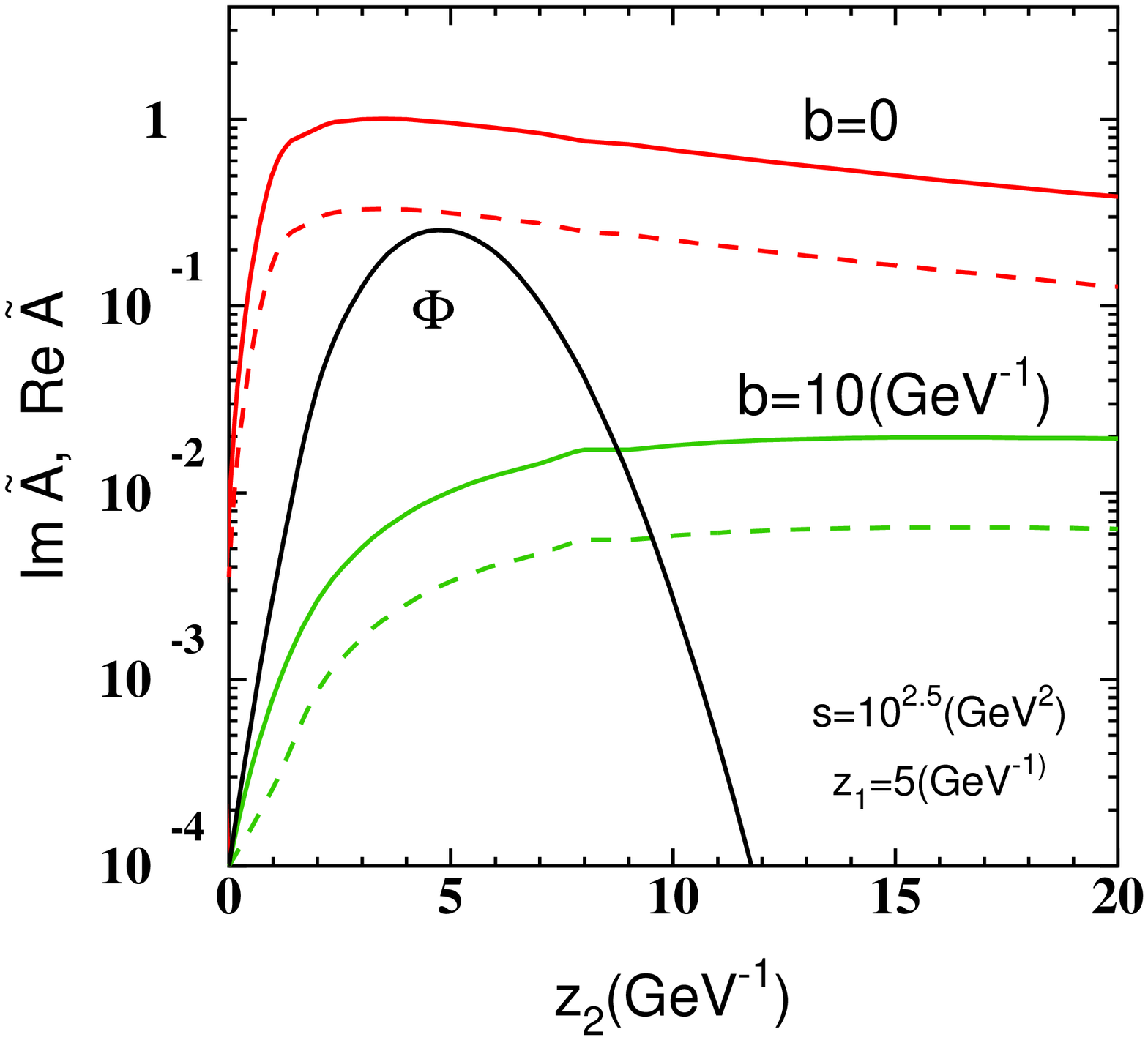,width=60mm,height=65mm}&~~~~~~~~~~~~~~~~&
\epsfig{file=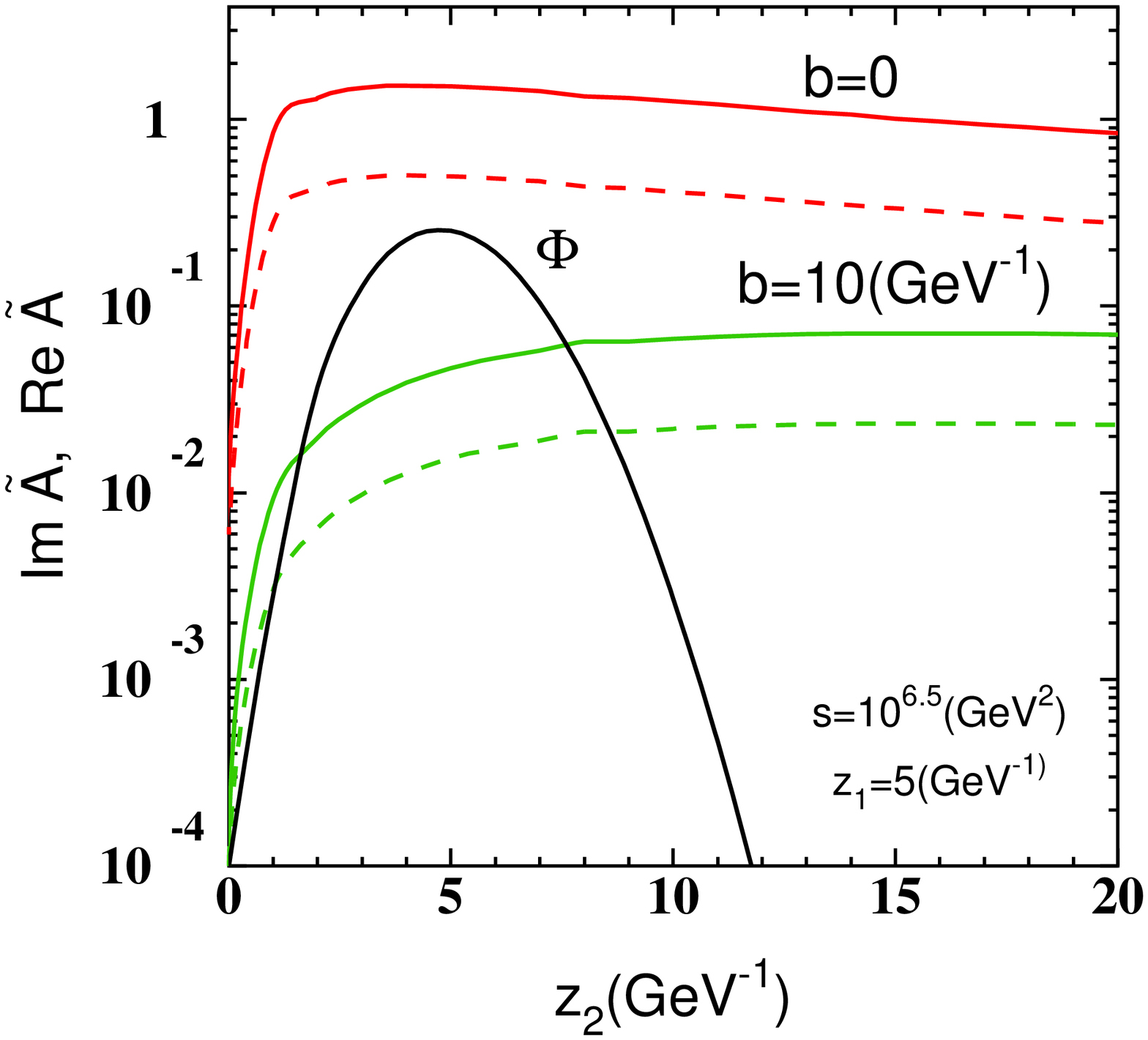,width=60mm,height=65mm}\\
\fig{am}-a & &\fig{am}-b\\
\end{tabular}
\caption{The behaviour of the real (dotted line) and the imaginary (solid line) parts of the scattering amplitude,
 and $\Phi\Lb z \Rb$ versus $z$ for different energies: $log_{10} s = 2.5$ (\fig{am}-a) and  $log_{10}s = 6.5$  (\fig{am}-b) and impact parameters ($b$).}
\label{am}
}

The first conclusion from our fit, is that we need to
find a new mechanism for diffractive processes.  Actually, we know the missing ingredient in $N=4$ SYM.
 In the routine Pomeron approach, a substantial part of diffraction relates to the diffractive production of large masses, which stem from the triple Pomeron interaction
 (see \fig{3P}-a). In our approach we do not have the triple graviton vertex.  Therefore, we believe that we need to find up to  what order of $\rho^n$ in the expansion,
 this vertex appears in $N=4$ SYM. It should be stressed that in $N=4$ SYM with small coupling (weakly interacting $N=4$ SYM), the diffractive dissociation due to the
 triple BFKL Pomeron plays an essential role.

\FIGURE[h]{
\begin{tabular}{c c}
\epsfig{file=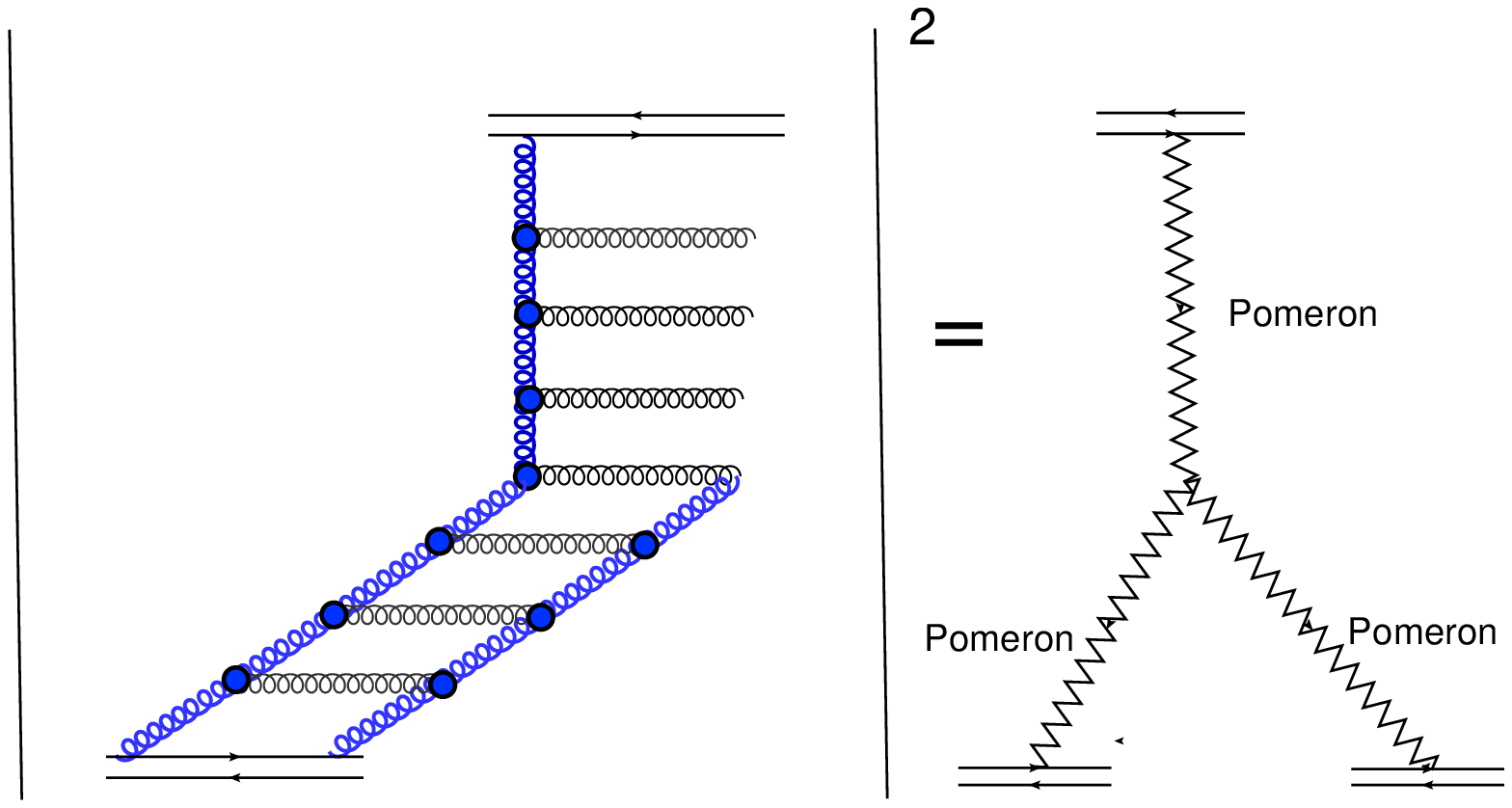,width=80mm}&\epsfig{file=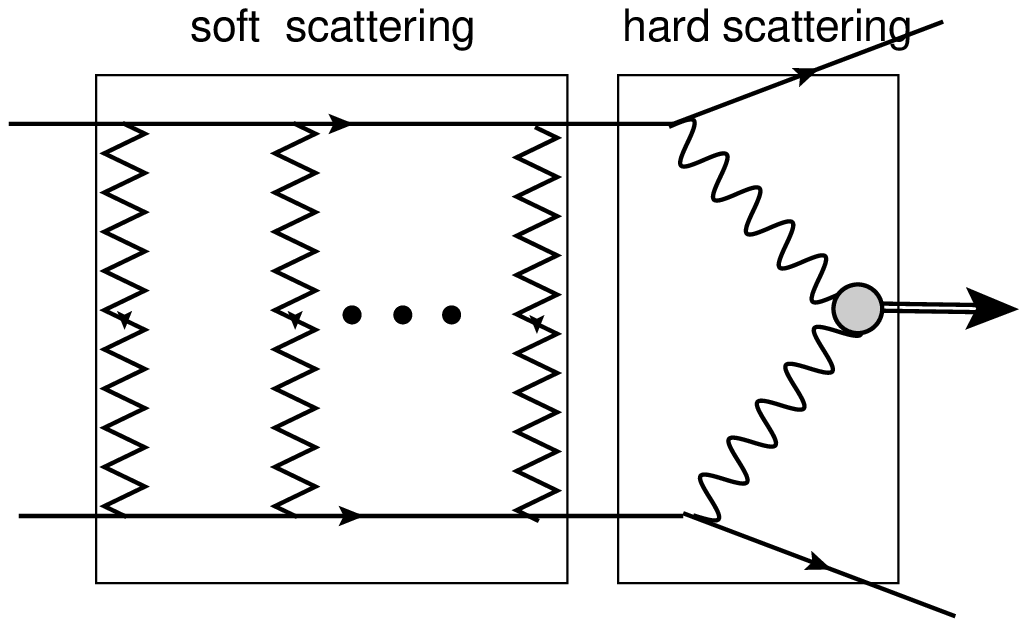,width=80mm,height=40mm}\\
\fig{3P}-a & \fig{3P}-b\\
\end{tabular}
\caption{The simple diagrams for large mass diffraction and its relationship with the triple Pomeron exchange in weakly coupled $N=4$ SYM (\protect\fig{3P}-a);
 and  for  central Higgs production (\protect\fig{3P}-b).
\label{3P} }
}

\subsection{Comparison with the LHC data}

Now let us consider the cross sections and other observables, within the LHC range of energies.
 As  mentioned above we failed to describe the cross sections for single and double diffractive dissociation,  obtaining extremely  small values for both ($\leq 2 mb$).
  However,  it  turns out  that the model can describe the data for the total, inelastic and elastic cross sections.
The comparison of our model with the LHC data is presented in Table 1. One can see that we predict the LHC data quite well, except for the value of the slope.
 In \fig{dsdt} we show the $t$ dependence of the elastic cross section  of our model, and the comparison of this distribution with the TOTEM experimental data.
 One can see that we overshoot the data, although the qualitative behaviour is reproduced quite well. We believe that to a large extent, our failure to describe 
 the data is correlated with a sufficiently small slope that we have (see Table 1).
 The value of the ratio $\mbox{Re} A/\mbox{Im} A$ that we obtain in the model, is larger than the value that uses TOTEM to extract the value of the total cross section 
($\mbox{Re} A/\mbox{Im} A$  = 0.14  \cite{REIM}), but the error coming from this difference, is well within the
experimental error for the TOTEM value of the total cross section, namely, $\sigma_{tot}=96.8 \,mb$.

\TABLE{

\begin{tabular}{|l |l|l|l|l|l|}

\hline \hline
W & $\sigma^{model}_{tot}$ &  $\sigma^{exp}_{tot}$ &  $\sigma^{model}_{el}$& $\sigma^{exp}_{el}$& Re /Im$^{model}$\\
\hline
7 TeV & 98.09 mb& TOTEM: 98.3 $\pm$0.2$^{st}$ $\pm$2,8$^{syst}$mb & 25.7 mb & TOTEM: 24.8$\pm 0.2^{st} \pm 1.2^{syst}$mb& 0.213\\
\hline

\end{tabular}

\begin{tabular}{|l|l|l|l|l|}

\hline
 W & $\sigma^{model}_{in}$& $\sigma^{exp}_{in}$&$B^{model}_{el}$& $B^{exp}_{el}$\\
\hline
7 TeV  & 72.4 mb & CMS:: 68.0$\pm 2^{syst} \pm 2.2^{lumi} \pm 4^{extrap}$ mb &18.3 $GeV^{-2}$&
TOTEM: 20.1$\pm 0.2^{st} \pm 0.3^{syst}\,GeV^{-2}$\\
 &  &  ATLAS: 69.4$\pm 2.4^{exp}  \pm 6.9^{extrap}$ mb & &\\
& &  ALICE: 72.7 $\pm 1.1^{model}  \pm 5.1^{extrap}$ mb & & \\
& & TOTEM: 73.5  $\pm 0.6^{st}  \pm 1.8^{syst}$ mb& & \\
\hline \hline
\end{tabular}

\caption{The comparison of the prediction of our model with the experimental data at W=7 TeV.}

}

\FIGURE[h]{
\centerline{\epsfig{file=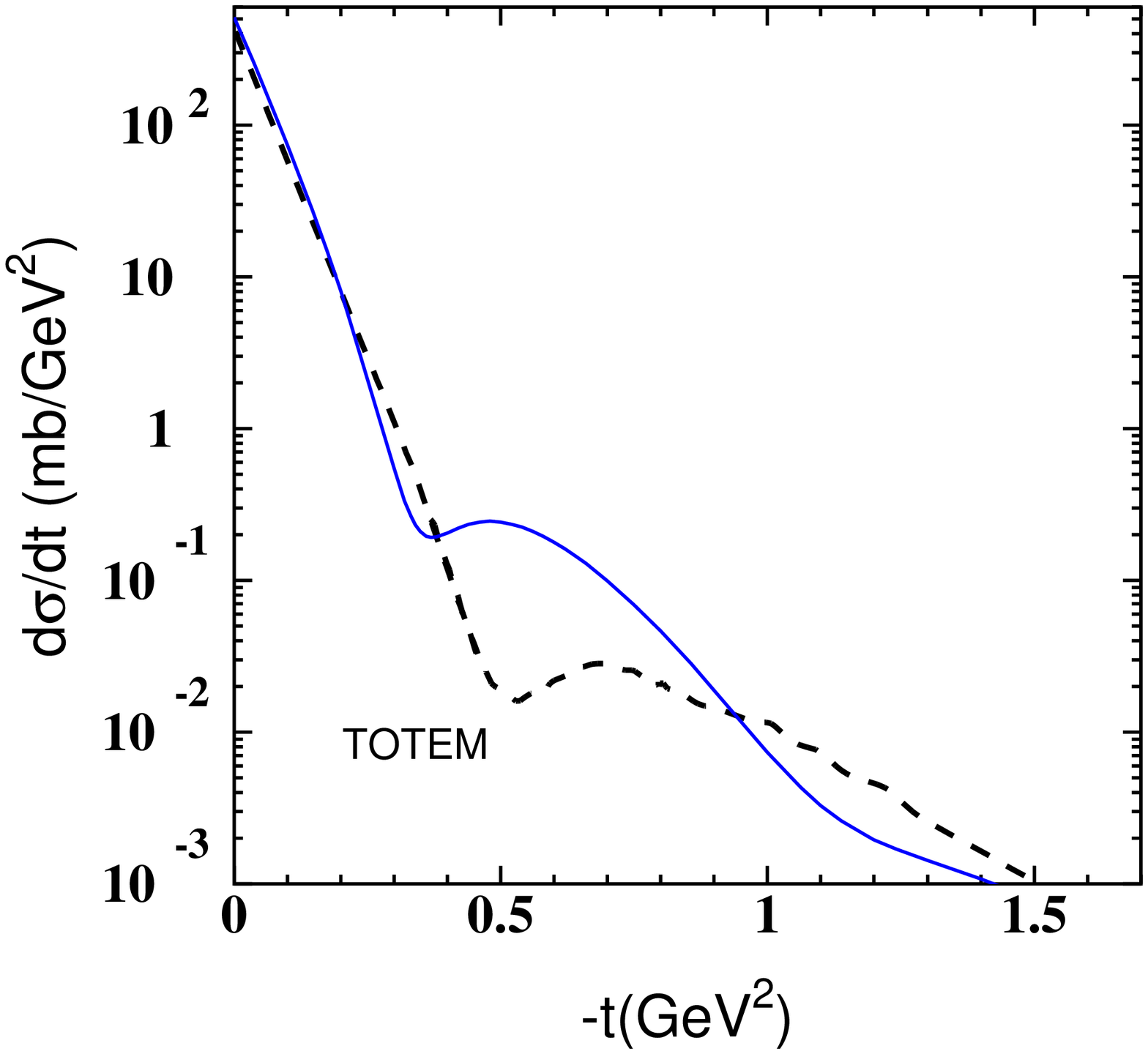,width=80mm}}
\caption{$d \sigma_{el}/d t $ of our model  (solid line), and the eye fit of the  TOTEM experimental data (\cite{TOTEM}) (dotted line).}
\label{dsdt} }~

\subsection{Survival probability of large rapidity gaps}

For a long time it has been known that  the cross section for processes with large rapidity gaps (for example central diffractive Higgs production (see \fig{3P}-b))
has to be multiplied by a factor  $S^2$ which is called the survival probability \cite{BJ,DOK,GLM1}.
 This factor stems from the possible interaction of the constituents of the projectile, with the target that should be forbidden to preserve the gap.
In other words, the constituents of the projectile could interact with the target in the initial or final state, suppressing the cross section for such processes 
(see \fig{3P}-b). The straitforward generalization of the well known formulae  \cite{BJ,DOK,GLM1}
leads to the following expression for the survival probability $S^2$:
\beq \label{S2}
S^2\,\,=\,\,\frac{\int d^2 b \int d z\, d z'\, \Phi\Lb z\Rb \Phi\Lb z'\Rb 
\, e^{ - \mbox{Im} A_\pom\Lb s,b; z, z'\Rb}\,A^2_{hard}\Lb b\Rb}{ \int d^2 b A^2_{hard}\Lb b\Rb}
\eeq

where $A_{hard}$ is the amplitude for Higgs production, in which the main contribution  stems from short distances.
One can see from \eq{S2} that the value for the survival probability depends on the impact parameter dependence,
 but not on the magnitude of the cross section. For simplicity we choose the exponential parametrization for $A^2_{hard}(b)$, namely,

\beq \label{B}
A^2_{hard}\,\,\propto\,\,\exp\Big( - \frac{b^2}{2 B}\Big)
\eeq

where $B$ is the slope of the hard differential cross section.
In \fig{dissl}  which is taken from Ref. \cite{HERA}, the experimental values of the slope for DIS diffractive production are plotted.

\FIGURE[h]{
\centerline{\epsfig{file=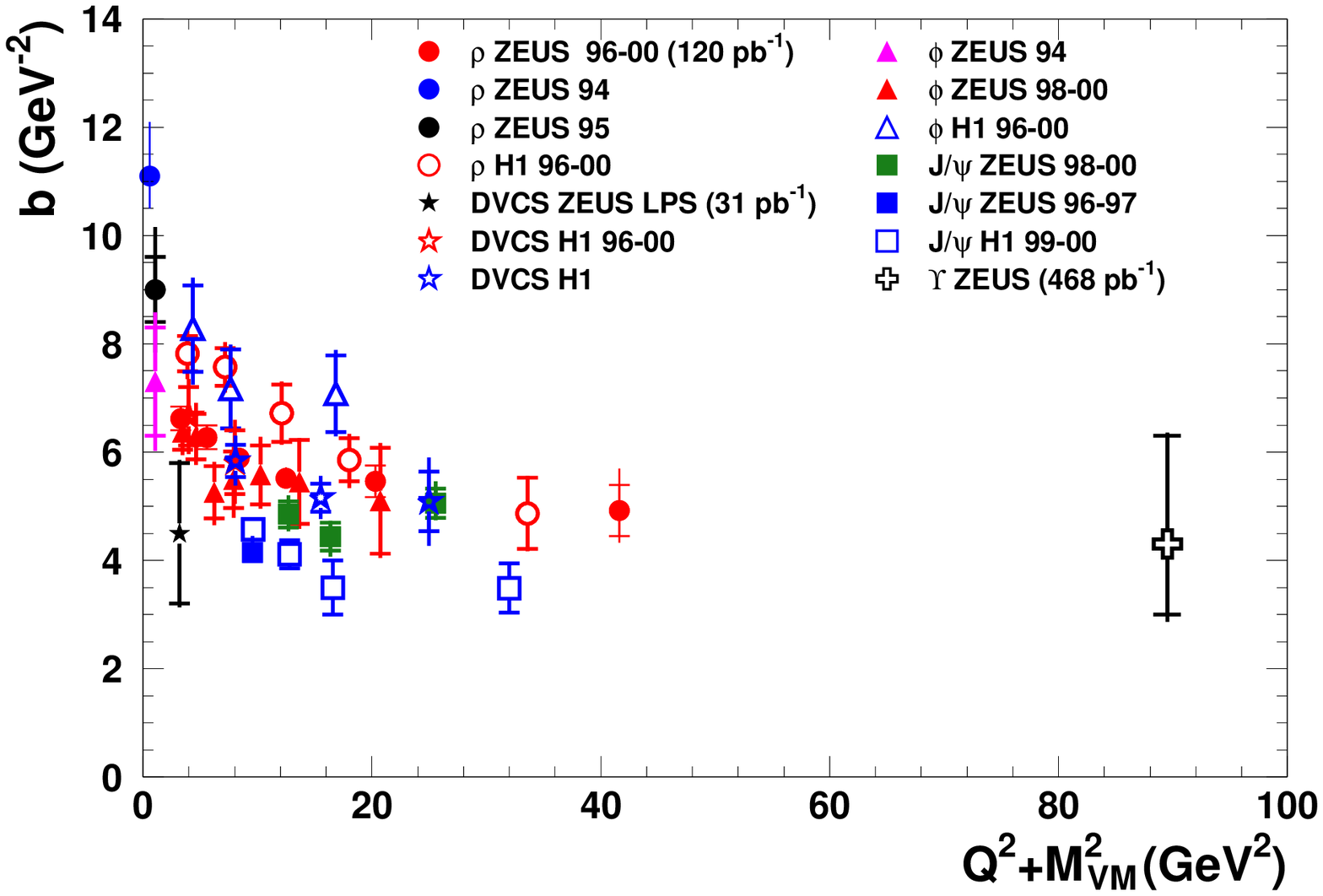,width=100mm}}
\caption{The dependence of the slope for DIS processes ($\sigma_{el}/d t \propto exp\Lb- B t\Rb$. The picture is taken from Ref.  \cite{HERA} .}
\label{dissl} }

From \fig{dissl} we can conclude that the average $B$ at large momentum scale (short distances),  is about $4 \div 5 \,GeV^{-2}$. We made estimates
 for $S^2$ for two values of $B$: $ B=4 \,GeV^{-2}$ and  $B = 5GeV^{-2}$,
These estimates lead to the following values

\bea \label{S2E}
 B = 5 GeV^{-2} &  \longrightarrow & TEVATRON:  S^2  = 0.126;\,\,\,\,
   LHC:   S^2 = 0.084;\nn\\
 B = 4 GeV^{-2} & \longrightarrow & TEVATRON:  S^2  = 0.106;\,\,\,\,
   LHC:   S^2 = 0.068
\eea

One can see that we obtain the value for $S^2$, which is $2 \div 3$ times larger than all previous estimates  \cite{KMRS,GLMSP}.

\section{Conclusions}

In this paper we showed that the model of $N=4$ SYM is able to describe the experimental data for the total, elastic and inelastic cross sections,
 and on the elastic slope in the range of energies from $W=20\, GeV$, up to the Tevatron energy. However, we failed to describe the cross sections for diffractive production.
  Our model gives these cross sections $\leq 2 mb$. The predictive power of the model is also rather limited.
The total inelastic and elastic cross sections are predicted by the model, but the model leads to the value of the slope which is smaller than the TOTEM data.
We  made a fit in which we include the LHC data in the fitting procedure, to check the stability of our model.
 This fit is shown by the dashed line in \fig{sig} and \fig{bel}. One can see that the difference is very small. Therefore, we can claim  that  we 
are  not able to describe both the elastic slope and the cross sections for diffractive production.

The parameters that come out from the fit: $g_s = 0.245$, $\rho = 0.797$ and $a = 0.23$ $GeV^{-2}$ are large for using  $N=4$ SYM,
 in the limit of low $\rho$.  Therefore  the next order correction (at least  of the order of $\rho^2 $)   are needed for a better understanding of the origin of
the process of diffractive production in $N=4$ SYM.

On the other hand, we cannot avoid  the feeling that $N=4$ SYM, in the approach of fitting a number of soft observables  with only three free parameters,
 is closely related to the theory of strong interactions at high energies.

We consider this paper as a call for a calculation up to the order of  $\rho^2$ in $N=4$ SYM, in which we can find both the source of large mass diffraction,
 and the correction to the elastic slope.

\begin{acknowledgments}

 This work was supported in part
by Fondecyt (Chile) grants 1090236 and 1100648.
\end{acknowledgments}


\begin{thebibliography}{99}

\bibitem{GLM}
 E.~Gotsman, E.~Levin and U.~Maor,
  Eur.\ Phys.\ J.\  C {\bf 71} (2011) 1553
  [arXiv:1010.5323 [hep-ph]].
\bibitem{KAP}
 A.~B.~Kaidalov and M.~G.~Poghosyan,
  arXiv:0909.5156 [hep-ph].

\bibitem{KMR}
 A.~D.~Martin, M.~G.~Ryskin and V.~A.~Khoze,
  arXiv:1110.1973 [hep-ph].
\bibitem{OST}
S.~Ostapchenko,
  Phys.\ Rev.\  D {\bf 83} (2011) 014018
  [arXiv:1010.1869 [hep-ph]].

\bibitem{ALICE}
M.~G.~Poghosyan,
  J.\ Phys.\ G G {\bf 38}, 124044 (2011)
  [arXiv:1109.4510 [hep-ex]].
\bibitem{ATLAS}
G.~Aad {\it et al.}  [ATLAS Collaboration],
  Nature Commun.\  {\bf 2} (2011) 463
  [arXiv:1104.0326 [hep-ex]].
\bibitem{CMS}
CMS Physics Analysis Summary:
Measurement of the inelastic pp cross section at √s = 7 TeV with the CMS detector", 2011/08/27./


\bibitem{TOTEM}
 F.~Ferro [TOTEM Collaboration],
  AIP Conf.\ Proc.\  {\bf 1350} (2011) 172;\,\,\,G.~Antchev {\it et al.}  [TOTEM Collaboration],
  Europhys.\ Lett.\  {\bf 96} (2011) 21002,
  {\bf 95} (2011) 41001
  [arXiv:1110.1385 [hep-ex]].

\bibitem{AdS-CFT}
 J.~M.~Maldacena,
  Adv.\ Theor.\ Math.\ Phys.\  {\bf 2} (1998) 231
  [Int.\ J.\ Theor.\ Phys.\  {\bf 38} (1999) 1113]
  [arXiv:hep-th/9711200];\,\,\,
S.~S.~Gubser, I.~R.~Klebanov and A.~M.~Polyakov,
  Phys.\ Lett.\  B {\bf 428} (1998) 105
  [arXiv:hep-th/9802109];\,\,\,
E.~Witten,
  Adv.\ Theor.\ Math.\ Phys.\  {\bf 2} (1998) 505
  [arXiv:hep-th/9803131].
\bibitem{LEPO1}
E.~Levin and I.~Potashnikova,
  JHEP {\bf 0906} (2009) 031
  [arXiv:0902.3122 [hep-ph]].
\bibitem{POST}
 J.~Polchinski and M.~J.~Strassler,
  JHEP {\bf 0305} (2003) 012
  [arXiv:hep-th/0209211];\,\,
  Phys.\ Rev.\ Lett.\  {\bf 88} (2002) 031601
  [arXiv:hep-th/0109174].
\bibitem{MHI}
Y.~Hatta, E.~Iancu and A.~H.~Mueller,
  JHEP {\bf 0801} (2008) 026
  [arXiv:0710.2148 [hep-th]].




\bibitem{BST1}
 R.~C.~Brower, J.~Polchinski, M.~J.~Strassler and C.~I.~Tan,
  JHEP {\bf 0712} (2007) 005
  [arXiv:hep-th/0603115].
\bibitem{BST2}
 R.~C.~Brower, M.~J.~Strassler and C.~I.~Tan,
  arXiv:0707.2408 [hep-th].
\bibitem{BST3}
 R.~C.~Brower, M.~J.~Strassler and C.~I.~Tan,
  JHEP {\bf 0806} (2008) 048
  [arXiv:0801.3002 [hep-th]].
  \bibitem{COCO}
 L.~Cornalba and M.~S.~Costa,
 Phys. Rev. {\bf D 78}, (2008) 09010,
  arXiv:0804.1562 [hep-ph];\,\,\,
  L.~Cornalba, M.~S.~Costa and J.~Penedones,
  JHEP {\bf 0806} (2008) 048
  [arXiv:0801.3002 [hep-th]];\,\,
  JHEP {\bf 0709} (2007) 037
  [arXiv:0707.0120 [hep-th]].

\bibitem{BEPI}
B.~Pire, C.~Roiesnel, L.~Szymanowski and S.~Wallon,
  Phys.\ Lett.\  B {\bf 670}, 84 (2008)
  [arXiv:0805.4346 [hep-ph]].
\bibitem{LMKS}
E.~Levin, J.~Miller, B.~Z.~Kopeliovich and I.~Schmidt,
JHEP {\bf 0902} (2009) 048;\,\,
  arXiv:0811.3586 [hep-ph].


\bibitem{BFKL}
 E. A. Kuraev, L. N. Lipatov, and F. S. Fadin, {\it  Sov. Phys.
JETP}
                {\bf 45}, 199 (1977); \,\,\,
Ya. Ya. Balitsky and L. N. Lipatov,
               {\it   Sov. J. Nucl. Phys.}\, {\bf 28}, 22 (1978).











\bibitem{FROI}
M.~Froissart,
{\it Phys.\, Rev.} \,  {\bf 123} (1961) 1053; \\
~A. ~Martin, {``Scattering Theory: Unitarity, Analitysity and Crossing."}
Lecture Notes in Physics, Springer-Verlag,  Berlin-Heidelberg-New-York,
1969.


\bibitem{KOWI}
A.~Kovner and U.~A.~Wiedemann,
  Phys.\ Lett.\ B {\bf 551}, 311 (2003)
  [hep-ph/0207335] and reference therein.


\bibitem{SOFTW}
 A.~Karch, E.~Katz, D.~T.~Son and M.~A.~Stephanov,
  Phys.\ Rev.\ D {\bf 74} (2006) 015005
  [hep-ph/0602229];\,\,\, H.~R.~Grigoryan and A.~V.~Radyushkin,
  Phys.\ Rev.\ D {\bf 76} (2007) 095007
  [arXiv:0706.1543 [hep-ph]].

\bibitem{HADPROP}
S.~J.~Brodsky and G.~F.~de Teramond,
  Phys.\ Lett.\ B {\bf 582} (2004) 211
  [hep-th/0310227]; \,\,\,R.~A.~Janik and R.~B.~Peschanski,
  Nucl.\ Phys.\ B {\bf 565} (2000) 193
  [hep-th/9907177];\,\,\, H.~R.~Grigoryan and A.~V.~Radyushkin,
  Phys.\ Rev.\ D {\bf 76} (2007) 095007,
  [arXiv:0706.1543 [hep-ph]];\,\,\,P.~Colangelo, F.~De Fazio, F.~Giannuzzi, F.~Jugeau and S.~Nicotri,
  Phys.\ Rev.\ D {\bf 78} (2008) 055009
  [arXiv:0807.1054 [hep-ph]];\,\,\, A.~Vega and I.~Schmidt,
  Phys.\ Rev.\ D {\bf 82} (2010) 115023
  [arXiv:1005.3000 [hep-ph]];\,\,\, A.~Vega, I.~Schmidt, T.~Gutsche and V.~E.~Lyubovitskij,
  Phys.\ Rev.\  D {\bf 83} (2011) 036001
  [arXiv:1010.2815 [hep-ph]].
\bibitem{ABCA}
  Z.~Abidin and C.~E.~Carlson,
  Phys.\ Rev.\  D {\bf 79} (2009) 115003
  [arXiv:0903.4818 [hep-ph]].
\bibitem{MATH}
Abramowitz, M. and Stegun, I. A. (Eds.). "Confluent Hypergeometric Functions." Ch. 13 in Handbook of Mathematical Functions with Formulas, Graphs, and Mathematical Tables, 9th printing. New York: Dover, pp. 503-515, 1972.

\bibitem{PDG}
K. Nakamura et al. (Particle Data Group), J. Phys. G 37, 075021 (2010)
\bibitem{REIM}
N.~A.~Amos {\it et al.}  [E710 Collaboration],
  Phys.\ Rev.\ Lett.\  {\bf 68} (1992) 2433.

\bibitem{BJ}
J.D. Bjorken, 
Phys.Rev. D47 (1993) 101.
\bibitem{DOK}
Yuri L. Dokshitzer,  Valery A. Khoze, T. Sjostrand, 
Phys.Lett. B274 (1992) 116.
\bibitem{GLM1}
E. Gotsman, E.M. Levin, U. Maor,
Phys.Lett. B309 (1993) 199.

\bibitem{KMRS}
 M.~G.~Ryskin, A.~D.~Martin, V.~A.~Khoze {\it et al.},
  J.\ Phys.\ G {\bf G36 } (2009)  093001;\,\,\,
  Eur.\ Phys.\ J.\  {\bf C60 } (2009)  265-272;\,\,
Eur. Phys. J. {\bf C54} (2008) 199 [arXiv:0710.2494 [hep-ph]],

\bibitem{GLMSP}
E. Gotsman, E.M. Levin, U. Maor,
  Phys.\ Lett.\  {\bf B438 } (1998)  229;\,\,Phys.\ Rev.\  {\bf D60 } (1999)  094011,
  [hep-ph/9902294];\,\,  E.~Gotsman, E.~Levin, U.~Maor {\it et al.},
    [hep-ph/0511060];\,\, 
  Eur.\ Phys.\ J.\ C\ {\bf 71} (2011) 1685;\,\,E.~Gotsman, H.~Kowalski, E.~Levin {\it et al.},
  Eur.\ Phys.\ J.\  {\bf C47 } (2006)  655;
\bibitem{HERA}
H.~Abramowicz {\it et al.}  [ZEUS Collaboration],
  {\it`` Measurement of the t dependence in exclusive photoproduction of Upsilon(1S) mesons at HERA,''}
  arXiv:1111.2133 [hep-ex].
\end{thebibliography}
\end{document}